# UPDATE ON THE VERIFICATION AND VALIDATION EFFORTS FOR THE STAKEHOLDER TOOL FOR ASSESSING RADIOACTIVE TRANSPORTATION

Harish Gadey, Caitlin Condon, Steven Maheras, Kacey McGee

Pacific Northwest National Laboratory, Richland, Washington, United States of America 99338

**ABSTRACT**

The United States Department of Energy (U.S. DOE) is planning for the transportation, storage, and disposal of spent nuclear fuel (SNF) and high-level radioactive waste (HLW) from commercial nuclear power plants and other U.S. DOE sites. The Stakeholder Tool for Assessing Radioactive Transportation (START) is a web-based, geospatial decision-support tool developed for evaluating routing options and other aspects of transporting SNF and HLW via barge, train, truck, and intermodal surface transport in the continental United States. The verification and validation (V&V) effort is intended to independently assess START to provide confidence in the ability of the tool to accurately provide intended outputs. The results selected for the V&V effort of the START code include those identified as crucial outputs by subject matter experts.

Outputs from START such as shape files and keyhole markup language (KML) files are analyzed using a geodesic computation using the WSG-84 ellipsoid model. Most of the V&V efforts are aimed towards examining and comparing the total length reported in the various files in the START tool. This work also focuses on the development of V&V methodologies for various outputs that could be replicated by the end user on a set of user-defined routes. Over 150 origin destination pairs were run as part of this effort to test the functionality of the START tool. In addition to presenting results using an independent geodesic computation, this work will provide a comparison of the total route lengths between START version 3.3 and the previous release of START (version 3.2.2).

**INTRODUCTION**

The United States Department of Energy's (DOE) Office of Nuclear Energy (DOE-NE) continues to plan for an integrated waste management system to transport, store, and dispose of spent nuclear fuel (SNF) and high-level radioactive waste (HLW) as part of the Integrated Waste Management program (IWM) [1]. This involves implementing a flexible waste management system incrementally to ensure safe and secure operations [1]. As part of the implementation, DOE continues to develop the Stakeholder Tool for Assessing Radioactive Transportation (START), a web-based decision-support tool for the transportation of SNF and HLW [2].

START is designed to provide route data, among other capabilities, allowing for analysis and decision making when planning for the transportation of SNF and HLW in the US [3]. The tool is designed as a web-based application using an ArcGIS server that enables the user to initially select an origin and destination. This is followed by allowing the user to pick the preferred mode of transportation (heavy haul truck, rail, barge to rail, etc.). It is worth noting that some of the modes might not be available at certain sites, depending on the on-site infrastructure. The routing criteria the user wishes to use for this route is then selected, some of the options include minimum distance, minimum population, and minimum travel time. The user then selects one of the two buffer zone distances (800 meters or 2500 meters). Other routing options, such as stops or avoiding unwanted roads by setting route barriers, are also implemented. The selection of the various aforementioned options allows for flexibility in creating custom user-defined routes.

START version 3.3 is the latest version of the tool that was made available in August 2022. Some of the output data the START tool provides includes route length, travel time, accident likelihood, buffer zone population (800 and 2500 meters), track class, and estimated incident-free dose.

Once processed, routes can be displayed in the START user interface (Figure 1). Along with the data available within the user interface, various other output files like the Esri Shapefile, keyhole markup language (KML) file, summary details file as well as route details file can also be downloaded for further analysis and reporting. Each of these output files provides unique information about the route.

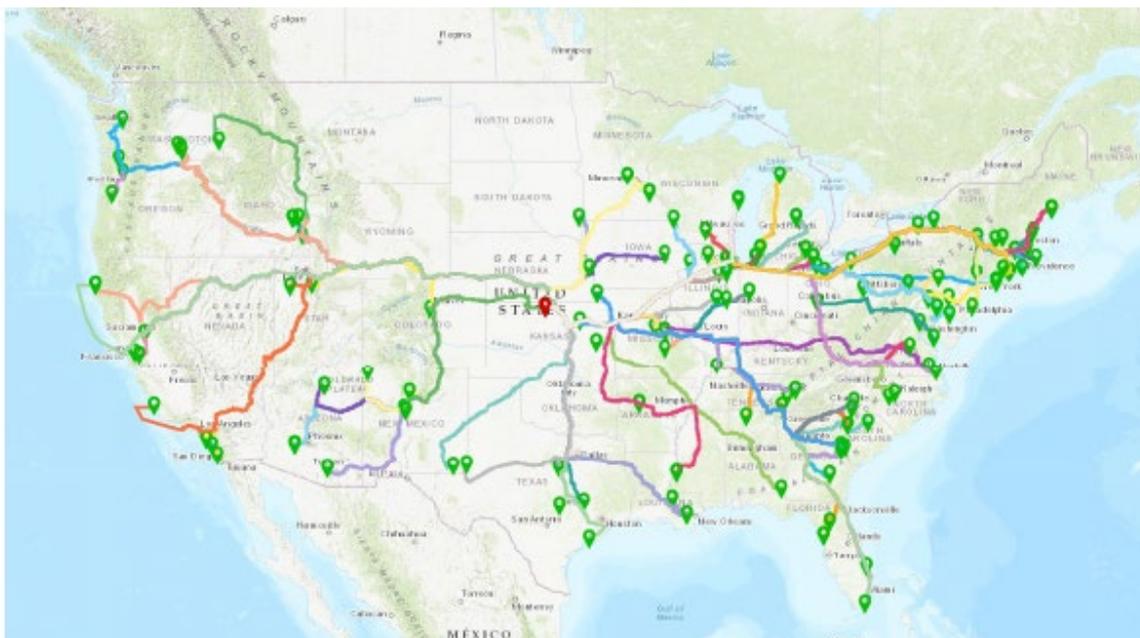

Figure 1. Example START V&V test routes

START uses an agile software development methodology in which the development team works alongside the verification and validation (V&V) team. This development model provides testing and quality control of features as they are developed and implemented in each new version. The features are tested, and the V&V team provides feedback to the development team. These suggestions are then implemented by the development team and are subsequently tested by the V&V team. This iterative development process between the development and the V&V team is called the agile software development process (Figure 2).

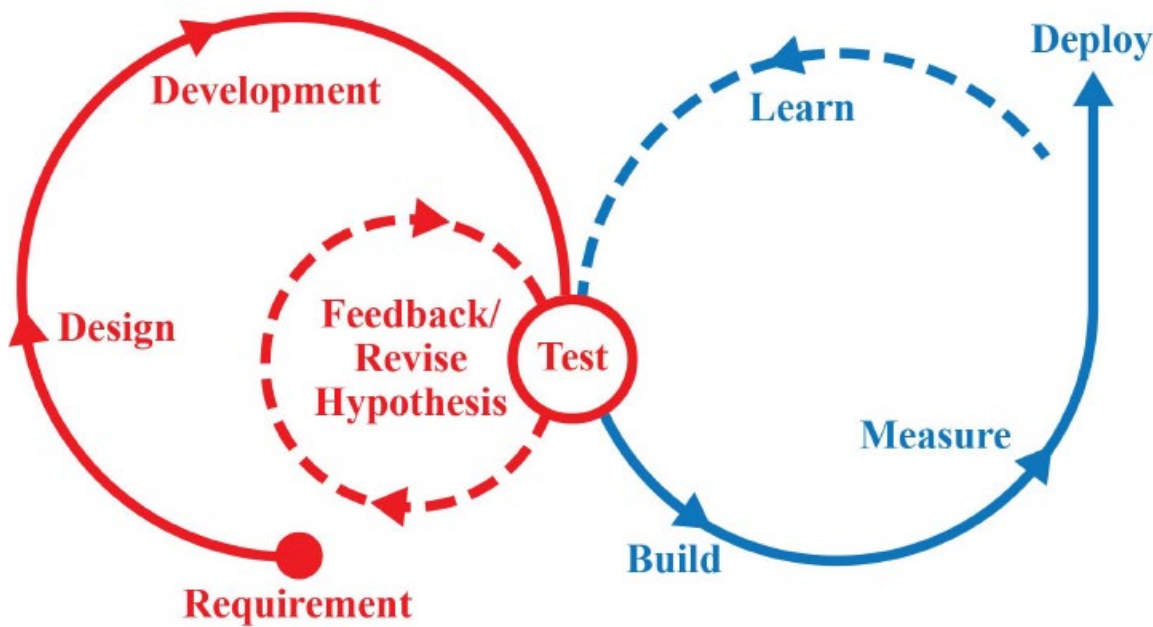

Figure 2. A conceptual figure describing the agile software development process

**START VERIFICATION AND VALIDATION**

Coordinate data of a route in START is provided by both the shapefile and KML file. While estimating the distance between coordinates on a curved surface such as the surface of the Earth, a geodesic

PNNL-SA-184469

computation is used to determine the shortest distance between the coordinates [4]. Due to the Earth's non-uniform ellipsoid shape, there are multiple ellipsoid models that could be used with varying degrees of accuracy depending on the geographic location. WGS-84 is a widely accepted ellipsoid model for computing geodesic distances on Earth and is the standard model used with KML files [5]. The mathematics used in the geodesic method as outlined by Karney [4] is implemented in the Python package GeoPy [6].

Most of the V&V test routes originate from a shutdown reactor, commercial operating reactor, or DOE facility site. The destination of all of the V&V test routes is currently set to the geographic center of the contiguous United States (Lebanon, KS). The test routes were run using the minimum distance criteria, with no added stops or barriers. The shapefile and KML output files generated from these test routes were used for the analysis presented in this work. In addition, test routes from previous V&V efforts can also be used to allow for a direct comparison between START versions.

The ESRI shapefile consists of a grouping of files containing data on geospatial features. Shapefiles cannot be opened with a standard text editor and require dedicated software such as ESRI's ArcGIS to be opened. As an output, the START-generated shapefile consists of a polyline geometry formed from multiple line segments (route segments). Each line segment is defined by dozens of data fields about the segment, containing information such as the population within a buffer zone, coordinates defining the segment, and the length of the segment. These files were analyzed using both ESRI's ArcGIS software for visualization as well as the PyShp package in Python for reading and parsing the files.

KML files are derived from the XML standard and are used specifically for geographic data [5]. KML was defined as an international standard in 2008 by the Open Geospatial Consortium and uses the WGS-84 ellipsoid model when projecting coordinates [5]. The START KML file output in version 3.3 consists of a single line string of coordinates providing a route overlay. The KML file lacks the data fields and segments provided in the shapefile. These files were analyzed using the quantum geographic information system (QGIS), an open-source geographic information system (GIS) tool for visualization, and a standard Python script for parsing.

In the Verification and Validation phase of START 3.3, comparisons were made between the reported lengths within the shapefile and geodesic calculated lengths using the shapefile coordinates. While it is possible to evaluate individual route segments, only the overall length of the route was compared. Additionally, the calculated geodesic lengths from the shapefile were compared to the calculated geodesic lengths of the KML file. Transportation mode-specific analysis was also performed for some of the output files.

Total Route Length Comparison (START 3.2.2 vs START 3.3)

The latest version of the START tool (version 3.3) was made available in August 2022. A comparison was performed for the total route length between START versions 3.2.2 and 3.3. Some of the major updates performed on the tool were migrating from the Idaho National Laboratory (INL) physical servers over to the Amazon Web Services (AWS) cloud server. In addition, updated GIS layers were also incorporated in the latest version of START. Since new data layers were being used, a total route length comparison was performed between the two versions. While performing the V&V tests, an initial test series consisting of 148 unique origin-destination pairs was created to test the functionality of the tool [7]. Out of a total of 148 unique origin-destination pairs, 15 of them failed to generate a route in START version 3.3. It is worth mentioning that the number of failed routes while using version 3.2.2 was 37. Therefore, a reduction in the number of failed routes was observed using the new version of START (3.3). Ninety-nine of the 148 routes were successfully generated in both versions of START. Three of the routes failed to successfully generate in START version 3.2.2 as well as START version 3.3. Table 1 provides data with respect to the change in distance observed for the routes between the two versions of START.

**Table 1.Total absolute route length change statistics for START 3.2.2 and START 3.3**

| Statistics | Value (%) |
|---|---|
| Minimum | 0.15 |
| Maximum | 42.12 |
| Mean | 6.05 |
| Median | 2.75 |
| Mode | 1.04 |
| Standard Deviation | 8.04 |

Shapefile Data Analysis

This analysis looked at the data provided in the shapefile. The total number of test routes was expanded from 148 to 169 routes to accommodate variable transportation modes for some of the origin-destination pairs. 150 out of the 169 routes were successfully generated and are used in this analysis. It is worth mentioning that some of the origin-destination pairs were run using multiple transportation modes to perform a robust analysis. The overall length of the summed segments showed good agreement with an independent measurement method (geodesic). The START shapefile provides a large amount of data broken down by route segment, including a listed segment length and geographic coordinates. The summation of the individual segment length data provided the total route length from the shapefile. This was compared to a summation of the calculated geodesic lengths using the shapefile coordinates.

The total route length data from START were in good agreement with the independent measurements from geodesic. On average, the absolute route lengths differed by 0.25% with a standard deviation of 0.11% (Figure 3). Routes were further evaluated individually by mode of transportation as shown in Figure 4. It is observed that the differences in total route length are relatively uniform across all modes of transportation.

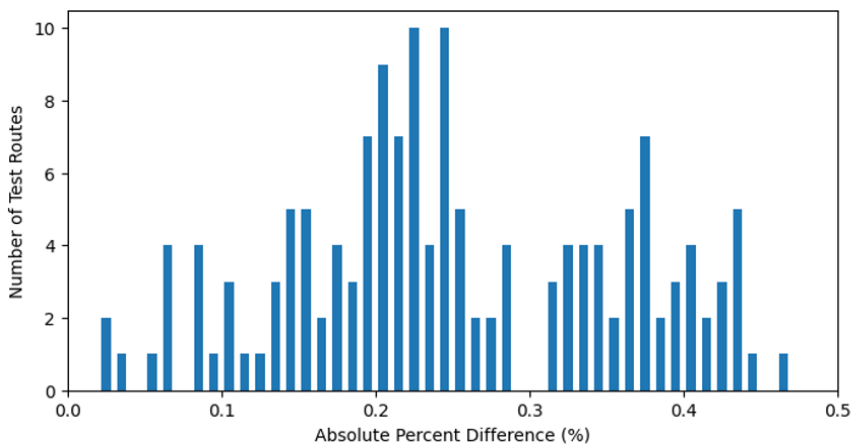

| Statistics | Value (%) |
|---|---|
| Minimum | 0.03 |
| Maximum | 0.36 |
| Mean | 0.25 |
| Median | 0.24 |
| Mode | 0.22 |
| Standard Dev. | 0.11 |

**Figure 3. Absolute percent differences between the START and geodesic reported data for the route length using the shapefiles along with some significant statistical results**

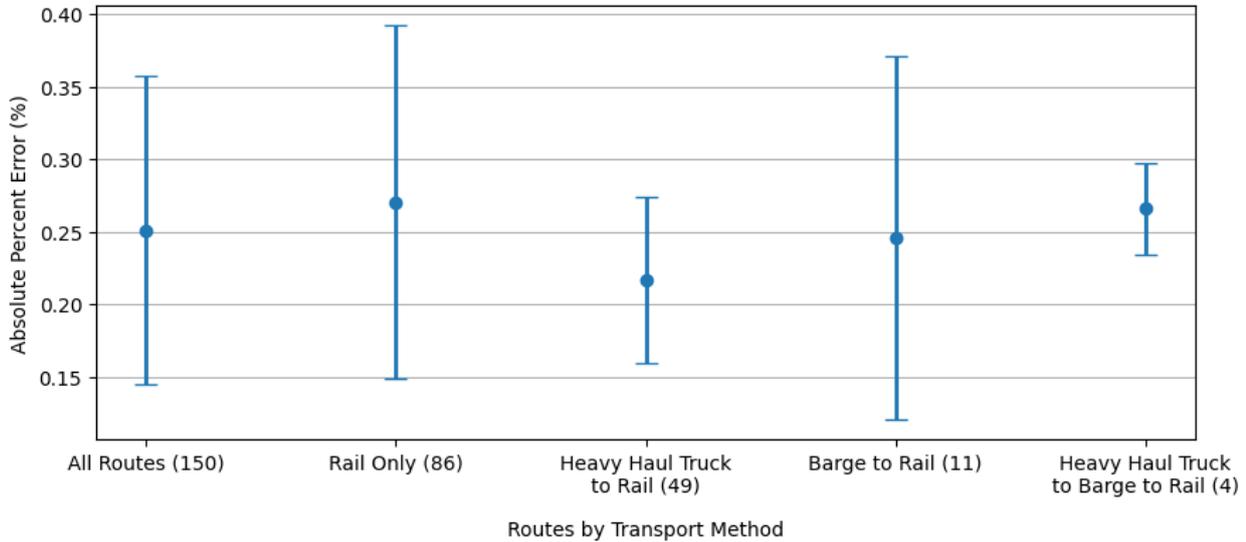

**Figure 4. Absolute percent differences between the START and geodesic reported data using shapefiles for route length employing various modes of transportation, the error bars represent one sigma standard deviation**

Shapefile and KML file Analysis

This section explores the differences in the KML and shapefile outputs using the calculated length from the coordinates provided in both file formats. In START 3.3, the KML file contains coordinate data to define the route generated but does not contain the data fields nor the route segments provided in the shapefile.

Evaluation of overall route length for each KML file was compared to the calculated length of the shapefile using the coordinates provided in each file. The calculation of the geodesic between adjacent coordinates were computed and summed to determine overall route lengths. Some modes of transportation were seen to exhibit a relatively large difference between the KML and shapefile files. Of the 150 test routes evaluated, only 13 routes had difference in length greater than 0.5%. Of the 15 routes evaluated using barge, 7 routes were seen to exhibit a difference greater than 0.5%, with one route presenting a difference of 12.44% between the two file types (KML and shapefile). Due to the relatively large differences in total distance observed in routes using barge transportation, these routes were additionally analyzed as their own data set. For routes using transportation modes other than barge, the absolute route lengths differed by 0.13% on average, with a standard deviation of 0.18% (Figure 5). In addition, all the routes were further evaluated individually by mode of transportation as shown in Figure 6. A separate analysis was performed for routes using the barge mode of transportation to gain additional insights.

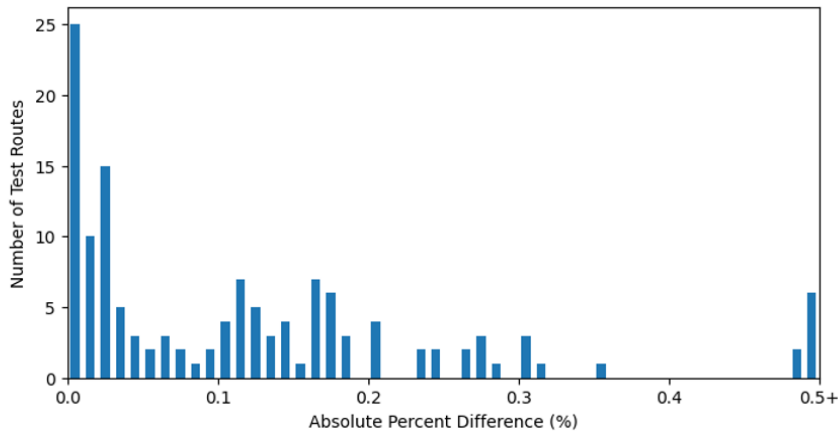

| Statistics | Value (%) |
|---|---|
| Minimum | 0.00 |
| Maximum | 1.26 |
| Mean | 0.13 |
| Median | 0.10 |
| Mode | 0.30 |
| Standard Dev. | 0.18 |

PNNL-SA-184469

**Figure 5. Absolute percent differences in route length using KML and shapefiles using geodesic measurements for all routes excluding those employing barge mode of transportation along with some significant statistical results**

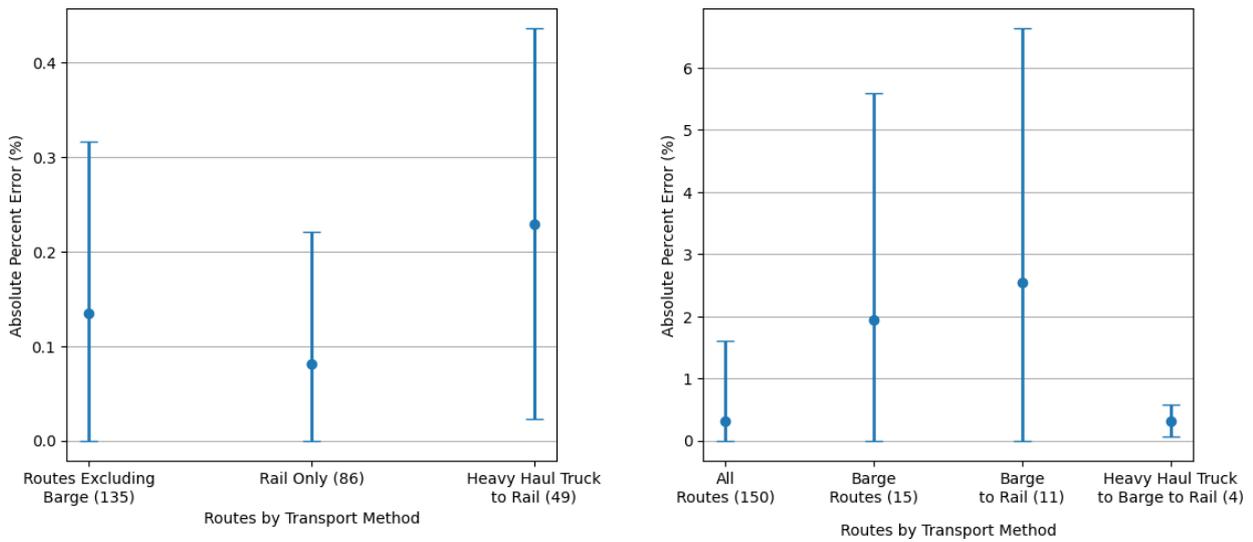

**Figure 6. Absolute percentage difference in route length between KML and shape files for all routes using geodesic measurements: excluding barge mode of transportation (left); using barge mode of transportation (right), the error bars represent one sigma standard deviation**

The routes using the barge transportation mode were seen to present the largest absolute differences in route length with a mean of 1.95% (SD 3.64%). While evaluating the barge data subset, it was determined that the largest differences were observed for routes starting with a barge. The routes using barges after using heavy haul truck (heavy haul truck to barge to rail) were observed to have similar differences to other modes of transportation.

For routes in which the largest differences were observed, more in-depth comparisons were performed between the KML output file and Shapefile output. The shapefile was converted to a KML file using ESRI's ArcMap software. Both files were then opened in QGIS for visualization. It was observed in Figure 7 that the output files defining the route have variable points of origin. The green route indicates the KML output file while the brown route indicates the shapefile output. This phenomenon was observed for several other test routes.

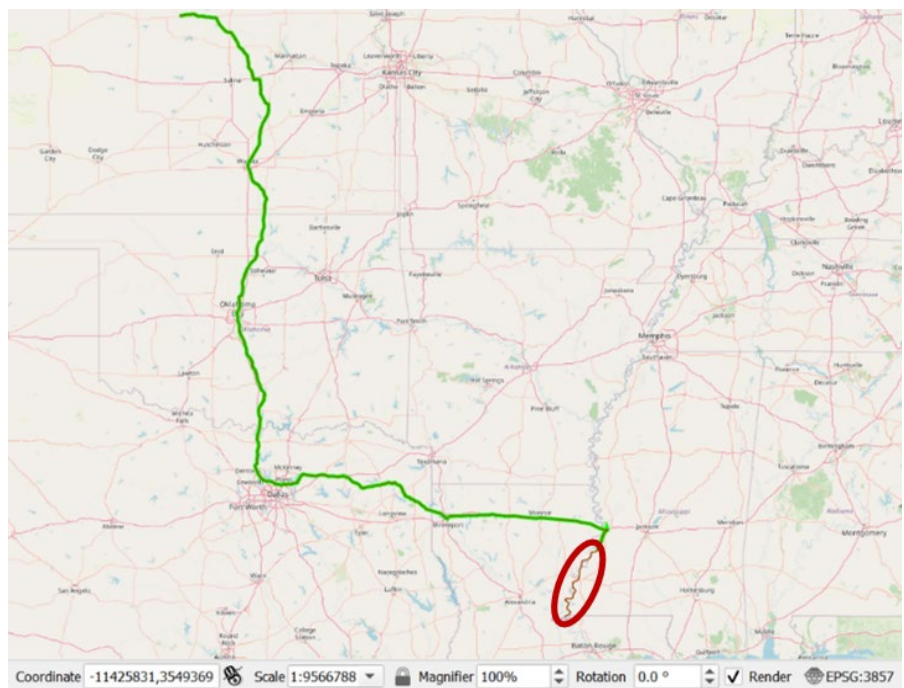

**Figure 7. START generated KML (green) and shapefile (brown) displayed in QGIS showing variation in routes as indicated by the red ellipse**

**FUTURE WORK**

Additional work is planned in the future to validate several other outputs from START version 3.3. It is initially planned to perform studies on the total route length using ESRI's ArcGIS and QGIS to perform individual validation of the total route lengths. This will be followed by estimating the population within the 800- and 2500-meter buffer zones for all test routes using QGIS. Once the population data is obtained, population densities are planned to be evaluated by dividing the population in the buffer zone by the total area under the buffer zone. Both the population and population density number can be validated from START reported data as well as from QGIS. Finally, radiation doses from incident free transport of SNF at a segment level are planned to be evaluated to verify the implementation of the methodology in the START tool.

**CONCLUSION**

START is a web-based decision-support tool intended to support planning and analysis for transport of SNF and HLW by DOE. Development of START follows an agile software development model with Verification and Validation (V&V) conducted alongside development. V&V provides testing and quality control of features implemented with each version update. In the latest START version 3.3, 150 test routes were created from various locations across the US and routed to Lebanon, KS (the geographic center of the contiguous United States).

The shapefile reported lengths showed good agreement compared to geodesic calculated lengths with an average difference of 0.25% (SD = 0.11%). When comparing the shapefile to the KML file computed length, an average difference of 0.32% (SD = 1.29%) was determined. An absolute difference of 0.13% (SD = 0.18%) is reported while excluding routes involving barge transportation. It was observed that KML files involving barge transport on average showed greater differences compared to the associated shapefile with an average difference of 1.95% (SD = 3.77%).

**ACKNOWLEDGEMENT**

Pacific Northwest National Laboratory is operated by Battelle Memorial Institute for the US Department of Energy under Contract No. DE-AC05-76RL01830. This work was supported by the US Department of Energy Office of Integrated Waste Management.

**ENDNOTE**

This is a technical paper that does not take into account contractual limitations or obligations under the Standard Contract for Disposal of Spent Nuclear Fuel and/or High-Level Radioactive Waste (Standard Contract) (10 CFR Part 961).To the extent discussions or recommendations in this report conflict with the provisions of the Standard Contract, the Standard Contract governs the obligations of the parties, and this report in no manner supersedes, overrides, or amends the Standard Contract. This report reflects technical work which could support future decision making by DOE. No inferences should be drawn from this report regarding future actions by DOE, which are limited both by the terms of the Standard Contract and Congressional appropriations for the Department to fulfill its obligations under the Nuclear Waste Policy Act including licensing and construction of a spent nuclear fuel repository.

PNNL-SA-184469